\documentclass[aps,prb,twocolumn,fleqn,superscriptaddress]{revtex4}  
\usepackage[english]{babel}
\usepackage{amsmath}
\usepackage{amssymb}
\usepackage{graphicx}
\usepackage{siunitx}
\usepackage[usenames, dvipsnames]{color}
\usepackage[super]{nth} 
\definecolor{color27}{rgb}{0.2, 0.6, 0.6}

\begin{document}

\title{Direct observation of plasma waves and dynamics \\induced by laser-accelerated electron beams}

\author{M. F. Gilljohann}
\affiliation{Ludwig-Maximilians-Universit\"at M\"unchen, Am Coulombwall 1, 85748 Garching, Germany}
\affiliation{Max Planck Institut f\"ur Quantenoptik, Hans-Kopfermann-Str.\ 1, Garching 85748, Germany}

\author{ H. Ding}
\affiliation{Ludwig-Maximilians-Universit\"at M\"unchen, Am Coulombwall 1, 85748 Garching, Germany}
\affiliation{Max Planck Institut f\"ur Quantenoptik, Hans-Kopfermann-Str.\ 1, Garching 85748, Germany}

\author{A. D\"opp}\email{andreas.doepp@physik.uni-muenchen.de}
\affiliation{Ludwig-Maximilians-Universit\"at M\"unchen, Am Coulombwall 1, 85748 Garching, Germany}
\affiliation{Max Planck Institut f\"ur Quantenoptik, Hans-Kopfermann-Str.\ 1, Garching 85748, Germany}

\author{J. G\"otzfried}
\affiliation{Ludwig-Maximilians-Universit\"at M\"unchen, Am Coulombwall 1, 85748 Garching, Germany}
\affiliation{Max Planck Institut f\"ur Quantenoptik, Hans-Kopfermann-Str.\ 1, Garching 85748, Germany}

\author{S. Schindler}
\affiliation{Ludwig-Maximilians-Universit\"at M\"unchen, Am Coulombwall 1, 85748 Garching, Germany}
\affiliation{Max Planck Institut f\"ur Quantenoptik, Hans-Kopfermann-Str.\ 1, Garching 85748, Germany}

\author{G. Schilling}
\affiliation{Ludwig-Maximilians-Universit\"at M\"unchen, Am Coulombwall 1, 85748 Garching, Germany}

\author{S. Corde}
\affiliation{LOA, ENSTA ParisTech - CNRS - \'Ecole Polytechnique - Universit\'e Paris-Saclay, 828 Boulevard des Mar\'echaux, 91762 Palaiseau Cedex, France }

\author{{A. Debus}}
\affiliation{Helmholtz-Zentrum Dresden - Rossendorf, Institute of Radiation Physics, Bautzner Landstrasse 400, 01328 Dresden, Germany}

\author{{T. Heinemann}}
\affiliation{Scottish Universities Physics Alliance, Department of Physics, University of Strathclyde, Glasgow G4 0NG, UK}
\affiliation{Deutsches Elektronen-Synchrotron DESY, D-22607 Hamburg, Germany}

\author{B. Hidding}
\affiliation{Scottish Universities Physics Alliance, Department of Physics, University of Strathclyde, Glasgow G4 0NG, UK}
\affiliation{Cockcroft Institute, Sci-Tech Daresbury, Keckwick Lane, Daresbury, Cheshire WA4 4AD, UK}

\author{S. M. Hooker}
\affiliation{John Adams Institute \& Department of Physics, Clarendon Laboratory, University of Oxford, Parks Road, Oxford OX1 3PU, United Kingdom}

\author{A. Irman}
\affiliation{Helmholtz-Zentrum Dresden - Rossendorf, Institute of Radiation Physics, Bautzner Landstrasse 400, 01328 Dresden, Germany}

\author{{O. Kononenko}}
\affiliation{LOA, ENSTA ParisTech - CNRS - \'Ecole Polytechnique - Universit\'e Paris-Saclay, 828 Boulevard des Mar\'echaux, 91762 Palaiseau Cedex, France }

\author{{T. Kurz}}
\affiliation{Helmholtz-Zentrum Dresden - Rossendorf, Institute of Radiation Physics, Bautzner Landstrasse 400, 01328 Dresden, Germany}

\author{A. Martinez de la Ossa}
\affiliation{Deutsches Elektronen-Synchrotron DESY, D-22607 Hamburg, Germany}

\author{U. Schramm}
\affiliation{Helmholtz-Zentrum Dresden - Rossendorf, Institute of Radiation Physics, Bautzner Landstrasse 400, 01328 Dresden, Germany}

\author{ S. Karsch}
\email{stefan.karsch@physik.uni-muenchen.de}
\affiliation{Ludwig-Maximilians-Universit\"at M\"unchen, Am Coulombwall 1, 85748 Garching, Germany}
\affiliation{Max Planck Institut f\"ur Quantenoptik, Hans-Kopfermann-Str.\ 1, Garching 85748, Germany}

\begin{abstract}
Plasma wakefield acceleration (PWFA) is a novel acceleration technique with promising prospects for both particle colliders and light sources. However, PWFA research has so far been limited to a few large-scale accelerator facilities world-wide. Here, we present first results on plasma wakefield generation using electron beams accelerated with a 100-TW-class Ti:Sa laser. Due to their ultrashort duration and high charge density, the laser-accelerated electron bunches are suitable to drive plasma waves at electron densities in the order of $\SI{e19}{\per\cubic\centi\metre}$. We capture the beam-induced plasma dynamics with femtosecond resolution using few-cycle optical probing and, in addition to the 
plasma wave itself, we observe a distinctive transverse ion motion in its trail. This previously unobserved phenomenon can be explained by the ponderomotive force of the plasma wave acting on the ions, resulting in a modulation of the plasma density over many picoseconds. Due to the scaling laws of plasma wakefield generation, results obtained at high plasma density using high-current laser-accelerated electron beams can be readily scaled to low-density systems. Laser-driven PWFA experiments can thus act as miniature models for their larger, conventional counterparts. Furthermore, our results pave the way towards a novel generation of laser-driven PWFA, which can potentially provide ultra-low emittance beams within a compact setup.
\end{abstract}

\maketitle

Over the past century, particle accelerators and colliders have been an essential tool to discover new physics. Electron accelerators based on radio frequency (RF) technology have pushed the frontier of high-energy physics to the \SI{100}{\giga\electronvolt}-level. However, to reach the TeV frontier, the limited acceleration gradient ($\lesssim \SI{100}{\mega\volt/\meter}$) of RF-technology means that tens of kilometers of acceleration length are required and such accelerators will eventually become too expensive to be built\cite{Tigner:2001im}. 
Accordingly, a number of alternative accelerator concepts have been explored over the last decades. One of the most promising is wakefield acceleration in plasmas \cite{Joshi:2003ur}, which relies on an intense particle or laser beam to excite a relativistic plasma wave with field strengths exceeding 100's of \si{\giga\volt/\meter}.\cite{Esarey:1996ui}

The concept of beam-driven plasma wakefield acceleration (PWFA) was developed in the 1980s\cite{Chen:1985wy,Chen:1986tb}. First experiments showing modest acceleration and the onset of self-focusing were performed shortly later at the Argonne National Laboratory \cite{Rosenzweig:1988kp,Rosenzweig:1990ig}. A major breakthrough was the observation of energy doubling of a \SI{42}{\giga\electronvolt} electron beam in an 85-cm-long PWFA at SLAC, which was reported in 2007\cite{Blumenfeld:2007ja}. More recent experiments also demonstrated an energy transfer efficiency exceeding $\SI{30}{\percent}$\cite{Litos:2015ke}, first high-energy positron acceleration \cite{Corde:2015ii} and GeV electron acceleration using proton-driven PWFA\cite{Adli:2018jz}. In the future, advanced injection methods are expected to provide ultra-low emittance electron beams \cite{Hidding:2012ep,Li:2013df,MartinezdelaOssa:2013hk,MartinezdelaOssa:2015es,Manahan:2017ky}, e.g. for compact free-electron lasers\cite{Maier:2012eq}, 
and proton-driven PWFA has the potential to accelerate electron beams to TeV-scale energies \cite{Caldwell:2009db}.

An important parameter to characterize electron beam drivers for PWFA is the peak charge density of the bunch, which is given by
\begin{equation}  
  \rho_\mathrm{b} = -e\cdot n_\mathrm{b} = \frac{Q}{(2\pi)^{3/2}\sigma_z\sigma_r^2} = \frac{I}{2\pi c \sigma_r^2}
  \label{nb}
\end{equation}
for a Gaussian beam. Here $e$ is the elementary charge, $n_\mathrm{b}$ the peak particle density, $Q$ denotes the beam charge, $I$ is the peak current, $\sigma_r$ is the root mean square (rms) transverse beam size and $\sigma_z$ is the rms bunch length. To exploit the multi-GV/m field gradients offered by the generation of nonlinear wakefields, $n_\mathrm{b}$ needs to be on the order of the plasma density $n_0$. In addition, the temporal bunch profile should be matched to the plasma wavelength 
\begin{equation}
  \lambda_\mathrm{p} =2 \pi c \sqrt{\frac{\epsilon_0 m_\mathrm{e}}{e^2 n_0}}\approx \SI{1}{\milli\metre} \times \sqrt{\frac{1}{n_0\left[\SI{e15}{\per \cubic \centi\metre}\right]}},
  \label{lambda_p}
\end{equation} 
with c the speed of light, $\epsilon_0$ the vacuum permittivity, $m_\mathrm{e}$ the electron mass and $e$ the elementary charge. 

The maximum accelerating field of a wakefield accelerator can be estimated by the cold wavebreaking field \cite{Esarey:2009ks} 
\begin{equation}
  E_0=\frac{2 \pi m_\mathrm{e} c^2}{e\lambda_\mathrm{p}}\approx \SI{3}{\giga\volt\per\metre}\times \sqrt{n_0[\SI{e15}{\per\cubic\centi\metre}]}
\end{equation}
and, accordingly, a PWFA needs to be operated at densities $\gtrsim \SI{e12}{\per\cubic\centi\meter}$ in order to generate higher accelerating fields than common RF-accelerators. But at the same time, meeting the above requirements to drive a wakefield becomes more challenging at higher plasma densities and currently only very few large-scale facilities worldwide are suitable to study PWFA and related plasma physics \cite{Joshi:2007ei,Hogan:2010ch}, typically at densities $n_0~\sim~\SI[parse-numbers = false]{10^{14}-10^{17}}{\per\cubic \centi\metre}$.  
Hence, numerical studies are often used to provide insight into the physics of PWFA, but the combination of micrometer plasma dynamics with meter-scale acceleration lengths requires simplified geometries and models to limit the computational costs \cite{Vay:2016gb}. 

\begin{figure*}[t]
  \centering
  \includegraphics[width=\textwidth]{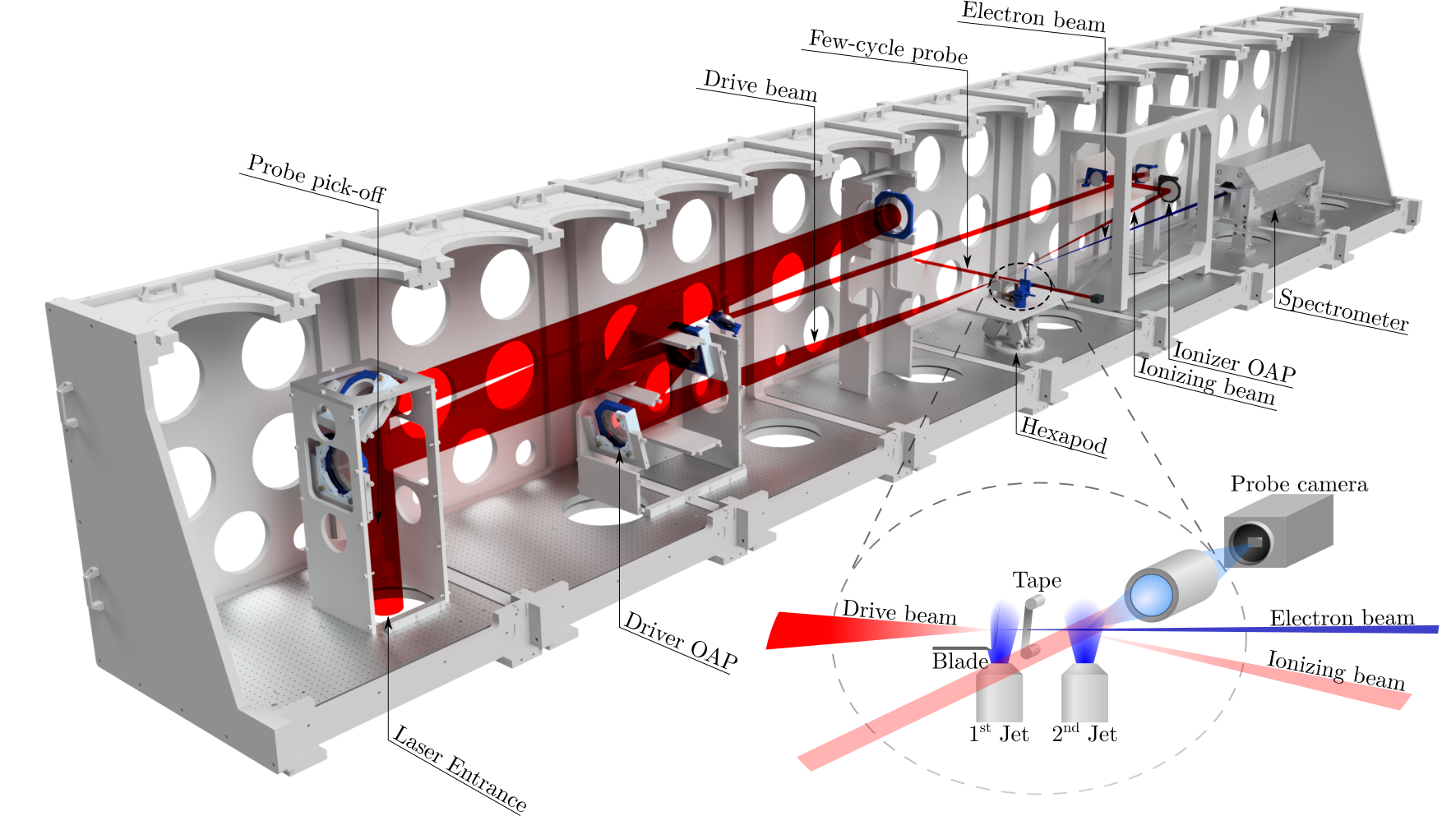}
  \caption{\textbf{Experimental setup.} The probe pulse is picked from the main beam at the chamber entrance and is coupled into a SPM-based broadening and compression setup outside of the vacuum chamber (not shown here). Meanwhile the drive beam (\SI{2}{\joule}, \SI{30}{\femto\second}) is delayed by $\sim\SI{20}{\nano\second}$ to accommodate for the additional delay of the few-cycle probe. The gas target and probe imaging setup are mounted on a hexapod stage in focus of the off-axis parabola (OAP). The profile of the laser-accelerated electron beam (indicated in blue) is measured with a scintillating screen (not shown here) mounted in front of the dipole magnet spectrometer. The ionizing pulse ($\sim\SI{60}{\milli\joule}$) is also picked from the main beam and focused using a second OAP at an angle of \SI{173}{\degree} to the drive beam. \textit{Bottom right:} Larger sketch of the target geometry, showing the two gas jets, the optional tape drive to block the laser and the three laser beams.}
  \label{fig:Setup}
\end{figure*}

Here we discuss a new experimental approach to study PWFA by using laser-wakefield-accelerated (LWFA)\cite{Malka:2008un,Esarey:2009ks,Hooker:2013jk} electrons as a plasma wave driver\cite{Hidding:2010es,Pae:2010du}. Due to their unprecedented peak currents and few-\si{\femto\second} duration\cite{Lundh:2011js} they allow the study of PWFA on much shorter spatial and temporal scales, corresponding to plasma densities in the \SI[parse-numbers=false]{10^{18}-10^{20}}{\per\cubic\centi\meter} regime and field gradients approaching \SI{100}{\giga\volt/\meter}, with commercially available \SI{100}{\tera\watt}-class Ti:Sa lasers as the primary driver.
As the physics of PWFA is completely scalable with the plasma density, depending only on the relative bunch density $n_\mathrm{b}/n_0$ and size $k_\mathrm{p}\sigma_{z|r} $ (with  $k_\mathrm{p}={2\pi}/\lambda_\mathrm{p}$), 
a LWFA-driven high-density PWFA can serve as a miniature model for large plasma accelerators such as FACET at SLAC\cite{Joshi:2018dn}, FLASHForward at DESY\cite{Aschikhin:2016fq} or AWAKE at CERN\cite{Gschwendtner:2016km}. It can thus provide a compact way to study physics related to beam-driven wakefield generation.

Beside its compactness, laser-driven PWFA offers several other advantages to its RF-driven counterparts. First, as they can be operated at densities exceeding \SI{e18}{\per\cubic\centi\meter}, it is possible to use shadowgraphy with few-cycle optical probes\cite{Schwab:2013if,Downer:2018ck} to study the interaction\footnote{With a central wavelength $\lambda_\mathrm{l} = \SI{800}{\nano\meter}$ the contrast of plasma waves in the shadowgram at densities below \SI{e18}{\per\cubic\centi\meter} is poor, due to the scaling of the refractive index $\eta = \sqrt{1-\lambda_\mathrm{l}^2/\lambda_\mathrm{p}^2}$. Lowering the density ($n_0 \propto \lambda_\mathrm{p}^{-2}$) even further e.g. by only two orders of magnitude to \SI{e16}{\per\cubic\centi\meter} the wavelength of the probe needs to be scaled accordingly by one order of magnitude to $\sim\SI{8}{\micro\meter}$, which is challenging on both laser and detector side.}. 
As these probes are usually derived from the same laser system, they are inherently synchronized to the laser-accelerated electron beam and can therefore provide snapshots of the plasma evolution with femtosecond jitter. Also, synchronized laser pulses can be used to provide accurately timed witness bunches, i.e. by techniques such as Trojan Horse injection\cite{Hidding:2012ep} for the production of low-emittance beams. Even the relatively large energy spread of the electron bunches typically generated by LWFA is beneficial for driving PWFA, because it suppresses beam hosing\cite{Vieira:2014cy}. 

So far, only indirect signs for a transition from LWFA to PWFA have been observed, based on either electron energy measurements\cite{MassonLaborde:2014he,Guillaume:2015gy}, pulse duration measurements \cite{Heigoldt:2015cd} or X-ray emission diagnostics\cite{Corde:2011}. First experiments dedicated to PWFA with laser-accelerated electron beams 
observed an electron deceleration signature\cite{Chou:2016ei} and electron beam focusing\cite{Kuschel:2016hy} in a second gas target. Here, we present the first direct and unambiguous observation of a plasma wave driven by laser-accelerated electrons using few-cycle shadowgraphy \cite{Schwab:2013if}. Furthermore, we present novel results on picosecond-timescale plasma ion dynamics behind the laser-generated electron beam driver, which demonstrate the capabilities of laser systems to advance PWFA research.

\subsection*{Experimental Methods}

\textit{Laser system.} The experiments were performed with the ATLAS laser at the Laboratory for Extreme Photonics, Garching. During the experiments, the Ti:Sa CPA system delivered 800 nm central wavelength laser pulses of \SI{28}{\femto\second} duration and \SI{2.5}{\joule} energy on target, corresponding to a peak power of \SI{84}{\tera\watt}.
 
\textit{Few-cycle shadowgraphy.} To obtain few-cycle probe pulses suitable for the shadowgraphy of plasma waves, a small part of the laser pulse ($\sim\SI{1}{\milli\joule}$) is coupled out before the focusing optics and sent into an Argon-filled hollow-core fiber. Self-phase modulation (SPM) inside the fiber leads to spectral broadening and allows to temporally compress the beam to below \SI{10}{\femto\second}, while its timing is adjusted with a delay stage (see the appendix for more details). 
It is sent through the target perpendicularly to the main pulse. The plane of interaction is imaged by a long working distance microscope objective (5x or 10x magnification, depending on the configuration) to form shadowgrams with a spatial resolution of approx.  \SI{2}{\micro\meter}. Due to the short pulse duration even rapidly moving structures like plasma waves can be resolved.
The measured diffraction signal directly reflects periodic modulations of the plasma density distribution, i.e. the laser- or beam-driven plasma wave. In the quasi-linear regime of wakefield acceleration, the periodicity of the plasma wave train is equal to the plasma wavelength $\lambda_\mathrm{p}$, which is \SIrange{10}{30}{\micro\metre} for densities of $n_0\sim \SI[parse-numbers = false]{10^{18}-10^{19}}{\per\cubic\centi\meter}$ (cf.~Eq.~\ref{lambda_p}). 
Meanwhile, the transverse size of the shadowgram depends not only on the wave's diameter, but also on the distance between plasma wave and the image plane, which is not precisely known due to the drive pulse's pointing jitter. By adding an optional Wollaston prism and a polarizer, the probe beam can also be used to implement an in-situ Nomarski-type interferometer to characterize the density of the plasma channel created by the drive beam.

\textit{Target configuration.} A 3D rendering of the setup in the vacuum chamber is shown in Fig.~\ref{fig:Setup}.
For laser wakefield acceleration, the horizontally polarized laser pulses are focused into a supersonic hydrogen gas jet target (hereafter referred to as the first jet) using a f/25 off-axis parabola, reaching an estimated peak normalized vector potential $a_0 = eA_0/m_\mathrm{e}c^2 = 1.6$ at focus. 
A subsequent hydrogen gas jet (hereafter referred to as the second jet) was installed downstream of the first jet, at variable distance and with independent flow control.
Optionally, a part of the main beam could be coupled out before the final focusing optics via a pick-off mirror and delay stage to provide an independently timed counter-propagating laser pulse (similar to Ref.
 \begingroup
    \romannumeral-`\x
    \setcitestyle{numbers}%
    \cite{Khrennikov:2015gxa}%
  \endgroup
  ) to ionize the second jet.

\textit{Laser wakefield accelerator.}
As a first jet, supersonic gas nozzles with \SI{3}{\milli\meter} and \SI{5}{\milli\meter} diameter were used. To facilitate electron injection, a silicon wafer was moved into the gas stream, leading to the formation of a shock front \cite{Schmid:2010ih,Buck:2013gs,Guillaume:2015dia}.
The jet was operated in a density range of \SIrange{3e18}{6e18}{\per\cubic\centi\meter}, which was in each specific configuration close to the threshold for self-injection. Shock-front injectors are usually operated at densities well below this threshold to generate monochromatic electron beams. Increasing the density leads to a higher energy spread but also substantially higher injected charge. 
This resulted in beams with up to \SI{900}{\pico\coulomb} in the energy range of \SIrange{25}{400}{\mega\electronvolt} at \SI{150}{\mega\electronvolt} central energy and down to \SI{0.6}{\milli\radian} FWHM divergence (see the appendix for representative electron spectra and Ref.\
\begingroup
    \romannumeral-`\x
    \setcitestyle{numbers}%
    \cite{Kurz:2018ji}%
  \endgroup~for details on the charge calibration). While the pulse duration was not directly measured in this experiment, previous bunch-length-measurements\cite{Heigoldt:2015cd,Lundh:2011js} suggest a duration of about 5 fs, corresponding to peak currents of up to \SI{170}{\kilo\ampere}.

\section*{Results}

\begin{figure*}[t]
  \centering
  \includegraphics*[width=\textwidth]{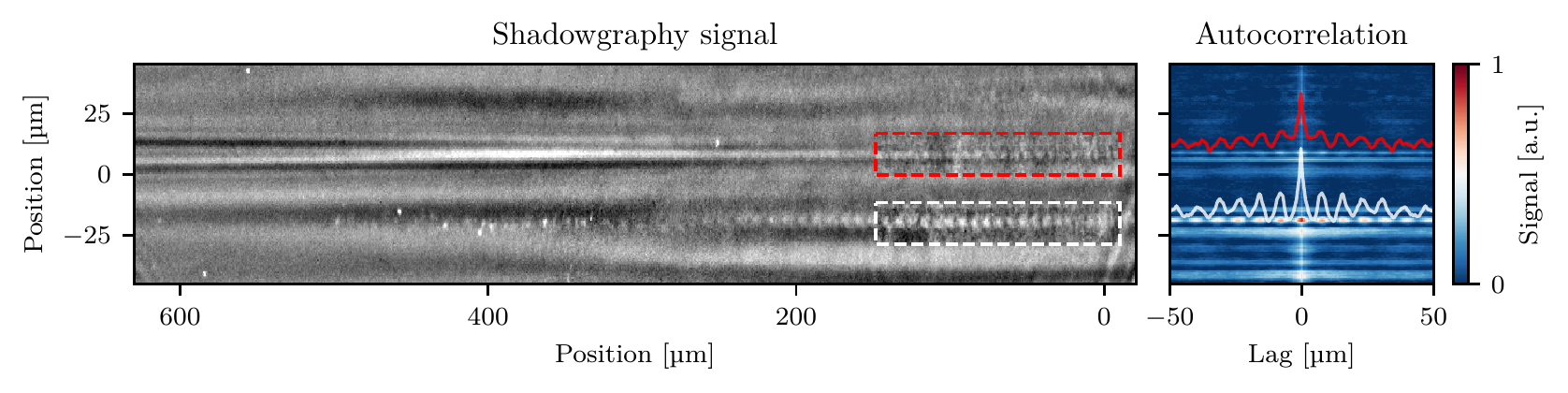}
  \caption{\textbf{Shadowgram of laser- and beam-driven plasma waves in the second gas jet.} \textit{Left:} Laser- and beam-driven plasma waves in the second gas jet (propagating to the right) after a free drift and spatial separation. Note the cone-like feature trailing only the upper plasma wave. \textit{Right:} Autocorrelation of each row of the signal in the interval of the marked plasma waves on the left. The red and white lineouts show the respective periodic signal modulations caused by the plasma waves.}
  \label{fig:LWFA_PWFA}
\end{figure*}

Here, we present the results of three experiments, each with a different configuration. In the \textit{first} setup we observed two plasma waves in the second jet, one of which has a distinct cone-like diffraction feature which we never observed for laser-driven plasma waves. This led to the assumption that this wave is driven by the electron beam from the first jet. To verify this hypothesis, we blocked the laser with a tape in the \textit{second} experimental configuration. When we pre-ionized the gas in the second jet we observed an unequivocally beam-driven plasma wave. Here, the diffraction feature is also visible. In a \textit{third} experiment we studied this cone feature, which turns out to be caused by the ion-motion.
A summary of the target parameters in each experiment can be found in Table~\ref{table:Configuration} in the appendix.

\subsubsection*{Observation of two plasma waves in a second gas target}

During LWFA, the electron beam is confined to the vicinity of the optical axis due to the transverse electrostatic wakefield\cite{Thaury:2015cg}. In this situation, the electron beam does not drive its own wave, but only affects the laser-driven wave via beam-loading\cite{Tzoufras:2008uj} until the laser depletes or the electron beam overtakes the laser. In both cases the laser will still perturb the beam-driven wave to a degree that is difficult to measure or predict.
In order to observe a purely beam-driven wave, one therefore needs to isolate the electron beam, i.e.~by blocking the laser with a foil\cite{Chou:2016ei}. However, scattering in the foil increases the electron bunch emittance and radius $\sigma_r$ after further propagation, which reduces its peak density $n_\mathrm{b}\propto \sigma_r^{-2}$. 

As an alternative, we exploited the fact that the electron beam pointing is not necessarily collinear to the laser axis. For instance, a slight pulse front tilt of the laser pulse can lead to skewed plasma wave fronts \cite{Popp:2010jl}. Hence, the laser and electron beam propagate at different angles in the space between both jets, leading to a spatial separation. In this \textit{first} experiment we generated a beam with  \SI{200}{\pico\coulomb} ($\sim \SI{40}{\kilo\ampere}$), \SI{0.6}{\milli\radian} FWHM divergence and a mean energy of \SI{150}{\mega\electronvolt} in the first jet. 
Indeed, as shown in Fig.~\ref{fig:LWFA_PWFA}, for most shots we observed two distinct plasma waves in the second jet, which is placed after a  \SI{3}{\milli\meter} vacuum gap behind the first jet. For the upper plasma wave we measure a wavelength of \SI[parse-numbers=false]{(7.6 \pm 0.1)}{\micro\meter}, for the lower one \SI[parse-numbers=false]{(7.8 \pm 0.1)}{\micro\meter}. The difference of \SI{2.6}{\percent} can be caused either a weak nonlinearity or a local difference of the plasma density $n_0= \SI[parse-numbers=false]{(1.9 \pm 0.1) \times 10^{19}}{\per\cubic\centi\meter}$. Accordingly, any laser contribution is expected to be weak, with a peak potential $a_0\lesssim 1$.

In principle it cannot be ruled out \textit{a priori} that both of these waves are driven by laser filaments. However, a marked difference in the morphology of both signals is the cone-like structure trailing one of the plasma waves. As will be discussed later, this feature is attributed to the dynamics of background plasma ions and a signature for electron-driven waves.

\subsubsection*{Observation of purely beam-driven plasma waves}
  
\begin{figure*}[t]
  \centering
\includegraphics[width=\textwidth]{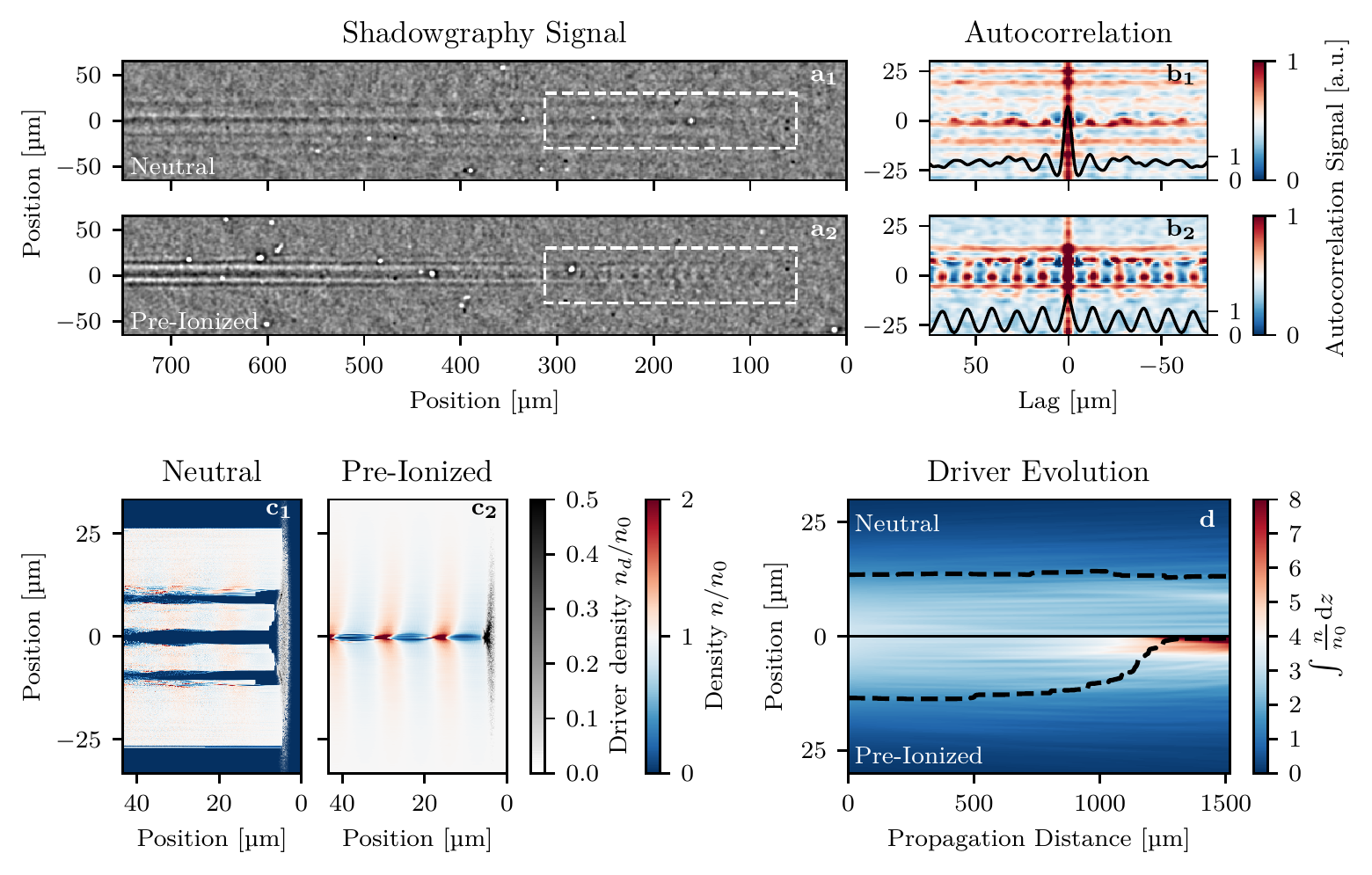}
\caption{\textbf{Electron-driven plasma waves with blocked laser.} \textit{Top:} \textbf{a}) Shadowgrams of the second jet for neutral and pre-ionized hydrogen gas. The drive bunch propagates from left to right.
\textbf{b}) Row-wise autocorrelations within the region of interest (dashed rectangle). The graphs in (b) are the horizontal lineouts at vertical zero.
The transverse modulation in the neutral case can only be attributed to ionization from the electron beam. The autocorrelation (b\textsubscript{1}) shows no indication of a longitudinal signal modulation that would be generated by a plasma wave. However, the pre-ionized case (a\textsubscript{2}) shows a weak, but visible periodical longitudinal modulation which is caused by a plasma wave driven by the electron beam. This modulation is clearly visible in the autocorrelation (b\textsubscript{2}). 
\textit{Bottom:} 
Full 3D-simulations of the interaction. \textbf{c}) Charge densities of the background plasma electrons (blue-red color scale) along with the driver (gray color scale). The regions in (c\textsubscript{1}) where the electron density is zero correspond to non-ionized gas.
\textbf{d}) Evolution of the transverse (longitudinally integrated) driver density for both cases, along with their half width at half maximum (HWHM, dashed line). The driver in the pre-ionized case self-focuses much faster and drives a stronger plasma wave, even with full blowout of the background electrons. In the neutral case the driver is not able to fully ionize the gas. See Fig.~\ref{suppl:Drivers} in the supplementary material for a close-up of the drivers.}
  \label{fig:PreionizedNeutral}
\end{figure*}

To verify that one of the plasma waves is really driven by an electron beam, we performed the \textit{second} experiment, where the setup was changed, such that the laser is blocked between the gas jets, with a \SI{15}{\micro\meter} thick 
Mylar tape acting as a plasma mirror\cite{Thaury:2007dg,Shaw:2016is}. As mentioned before, the foil defocusses the electron beam. In our measurements the divergence increased by a factor $\alpha = 2.7 \, \pm \, 1.5$, which results in a decrease of the wave amplitude by a factor of $\alpha^{-2}$.
This effect is minimized for small jet separations, yet, due mechanical constraints, the tape and the second jet were placed \SI{2}{\milli\meter} and \SI{10}{\milli\meter} respectively behind the end of the first jet.

In this configuration, the LWFA produced \SI{900}{\pico\coulomb}-electron beams.  Fig.~\ref{fig:PreionizedNeutral}a\textsubscript{1} shows that this bunch causes a transverse diffraction pattern in the shadowgram of the second jet, which indicates that the neutral gas was at least weakly ionized by the electron beam. However, there is no visible sign of a plasma wave and the autocorrelation of the data (Fig.~\ref{fig:PreionizedNeutral}b\textsubscript{1}) shows no obvious periodic features in longitudinal direction indicative of a plasma wave. This is likely the result of the missing pre-ionization by the laser and the fact that the foil-induced defocussing prevents the beam from becoming dense enough for causing more than weak ionization.
Ionization only occurs when the transverse electrostatic fields of the bunch exceed the field ionization threshold, which is about \SI{25}{\giga\volt\per\meter} for an ionization probability of \SI{1}{\percent} per fs in atomic hydrogen\cite{Bruhwiler:2003ez}. 
Hence, the head of the bunch, in front of the ionization, does not contribute to the wave generation.
Furthermore, in radial direction, the fields are zero in the center and reach a maximum at $\sigma_r$, which leads to an annular-shaped ionization trace. 

To overcome this problem, a counter-propagating pulse was used to pre-ionize the gas several picoseconds before the arrival of the electron beam. The ionization pulse had an energy of $\sim \SI{60}{\milli\joule}$ and intercepted the electron bunch at an angle of \SI{173}{\degree} to the driver axis.

In this case, the shadowgram in Fig.~\ref{fig:PreionizedNeutral}a\textsubscript{2} along with the autocorrelation (Fig.~\ref{fig:PreionizedNeutral}b\textsubscript{2}) shows a periodical longitudinal modulation at the plasma wavelength. Since the laser driver from the first jet was blocked by the tape this unequivocally is an isolated, purely electron-driven plasma wave. We measured a plasma wavelength of \SI[parse-numbers=false]{(13.6 \pm 0.3)}{\micro\meter}, which is in accordance with the plasma wavelength of \SI[parse-numbers=false]{(13.2 \pm 0.2)}{\micro\meter} (equivalent to $n_0=\SI[parse-numbers=false]{(6.2\pm0.2)\times 10^{18}}{\per\cubic\centi\meter}$)  from measurements without tape in otherwise identical conditions.  Note that the shadowgram shows a similar diffraction feature as observed behind one of the plasma waves in Fig.~\ref{fig:LWFA_PWFA}.

To verify our interpretation of the results, we performed full-3D particle-in-cell simulations using \textsc{Osiris} 4.4, with and without pre-ionization. The bunch exiting the first jet was measured to contain a total charge of \SI{900}{\pico\coulomb}, of which \SI{ 550}{\pico\coulomb} were transmitted through the second jet (see Fig.~\ref{suppl:ElectronSpectra} in the appendix). Half of the spectrum was detected in a low-energy (and/or highly divergent) background, which is unlikely to contribute significantly to the plasma wave generation. Thus, for the simulations only the  bunch charge between \SIrange{100}{350}{\mega\electronvolt} was considered, which amounts to \SI{300}{\pico\coulomb}.
The transverse size was calculated to $\sigma_r = \SI{11.8}{\micro\meter}$ from the average measured divergence. The spatially correlated momenta in the simulations were initialized according to the free drift with a divergence of \SI{1.7}{\milli\radian} and a temperature of \SI{40}{\kilo\electronvolt}. The temporal length was assumed to be \SI{5}{\femto\second} FWHM, which corresponds to a peak current of \SI{56}{\kilo\ampere}. 
The moving simulation box has a size of $(x \times y\times z)=(60\times60\times 20)\cdot k_\mathrm{p}^{-3}$ at a resolution of $\Delta x = \Delta y = \Delta z = 0.05k_\mathrm{p}^{-1} \simeq \SI{0.1}{\micro\metre}$ (with $n_0=\SI{6.4e18}{\per\cubic\centi\meter}$), and each cell is initialized with one electron macro particle. For simulations with an initially neutral gas, \textsc{Osiris} employs a field ionization model\cite{Bruhwiler:2003ez} to calculate ionization probabilities.

The simulation results are shown in Fig.~\ref{fig:PreionizedNeutral}c-d. We observe that the driver alone is not able to ionize the gas over its full extent (Fig.~\ref{fig:PreionizedNeutral}c\textsubscript{1}) and self-focuses much less than in the pre-ionized case (Fig.~\ref{fig:PreionizedNeutral}c\textsubscript{2} and \ref{suppl:Drivers} in the supplementary material for close-ups of the drivers). More specifically, it evolves into a funnel-like shape and ionizes two rings, which resembles the observed diffraction structure in Fig.~\ref{fig:PreionizedNeutral}a\textsubscript{1}. In contrast, in the pre-ionized case, the driver self-focuses much more strongly and in turn drives a higher amplitude plasma wave. 
The simulation also predicts that the tail can drive a few $\si{\micro\meter}$-radius plasma wave in full blow-out of the background electrons. However, such a small region would induce a weak phase-shift compared to the larger linear plasma wave and is therefore not observable in the shadowgraphy. 

To conclude, in this experiment we have observed beam-driven plasma waves at densities of $\sim \SI{e19}{\per\cubic\centi\metre}$ for the first time and, due to the tape, we can rule out any influence of the laser. 

\subsubsection*{Observation of ponderomotive ion channel formation}

\begin{figure*}[t]
  \centering
  \includegraphics*[]{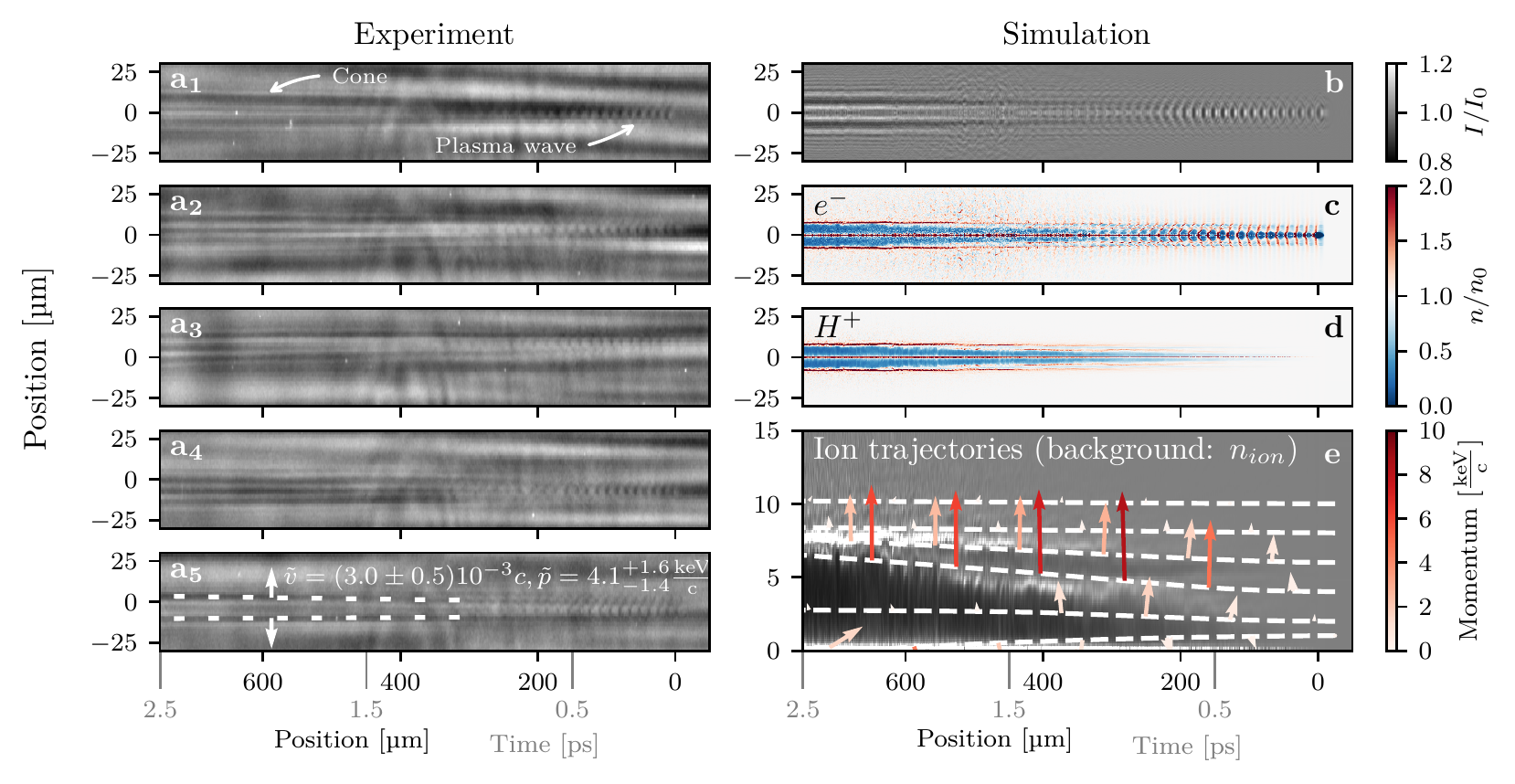}
  \caption{\textbf{Ion-channel formation from a plasma wakefield.} \textit{Left:} (\textbf{a}) Raw shadowgrams showing electron-driven plasma waves (propagating to the right) and their trailing ion channels for 5 consecutive shots. The dashed lines in the lower shadowgram exemplarily show the maxima of the ion distribution (via the electron distribution), the radial velocity of the maxima $\tilde{v}$ and the momentum of an ion with $\tilde{p} = m_\mathrm{i} \tilde{v}$. \textit{Right: } Corresponding particle-in-cell simulations and synthetic shadowgram (\textbf{b}). The electron (\textbf{c}) and ion densities (\textbf{d}) clearly show quasi-neutrality after several plasma wave periods. The channel in the synthetic shadowgram is in excellent agreement with the measured ones. The ion trajectories (\textbf{e}) with arrows indicating the instantaneous momenta, show that ions close to the symmetry axis are accelerated towards the axis, while ions with $r_0 \gtrapprox 2.5 k_\mathrm{p}^{-1}$ are accelerated in away from it.}
  \label{fig:IonChannels}
\end{figure*}

Beside the observation of a periodic intensity modulation from to the plasma wave, the shadowgrams also frequently show an unexpected, cone-like feature. So far, we have observed that this feature formed in most cases in the second jet without foil and always with foil (see Fig.~\ref{fig:PreionizedNeutral}a\textsubscript{2} and supplemental figures \ref{suppl:LWFA_PWFA} and \ref{suppl:PWFA} for a full field of view). In contrast, the feature is not present in any of the first-jet LWFA shadowgrams that we acquired and it is also absent in the second jet when there is no electron beam generated in the first jet. Therefore, we can conclude that the cone is indeed a distinguishing feature of electron-driven plasma waves, at least for our experimental conditions.

To further investigate this effect, we have performed a \textit{third} experiment that concentrates on the features of the cone. The large field of view of the shadowgraphy diagnostic allows to study the evolution on a picosecond timescale. 
In order to spoil the electron driver as little as possible, we removed the tape and moved the jets closer to each other. The configuration is similar to the first experiment, but with slightly increased separation and almost twice the density in the first jet (cf. Tab.~\ref{table:Configuration}). This leads to more than 2.5-times the beam charge (\SI{520}{\pico\coulomb}, $\sim\SI{100}{\kilo\ampere}$) and less transmitted laser energy. Accordingly, we observed only one plasma wave, always accompanied by a cone.
As shown in Fig.~\ref{fig:IonChannels}a and supplemental figures, its origin is located close to the tail of plasma wave, starting after a few hundred femtoseconds, and it persists at least out to $\SI{50}{\pico \second}$, as confirmed by varying the probe pulse delay. We measure a half opening angle $\alpha = \SI[parse-numbers=false]{(3.0 \pm 0.5)}{\milli\radian}$ of the cone in this specific configuration.

To our knowledge, no similar observation has been reported for either LWFA or PWFA and the origin of the diffraction cone was initially unclear. Assuming a mostly perpendicular motion, a transverse (group) velocity of $v_{\perp}=\SI{0.0017}{c}$ can be inferred from the opening angle. If the ion background was static and this feature arose only from electron motion, the velocities would be far too low to sustain a charge separation and the restoring forces would lead to plasma oscillations. Yet the latter are not observed and the feature has to be associated with ion motion.

We therefore performed PIC simulations with a mobile ion background. In order to cover the several picosecond-long experimental observation window, we assumed a symmetric beam driver and performed simulations in cylindrical coordinates. The drive bunch parameters were deduced from reference shots with the second jet switched off, i.e. \SI{520}{\pico\coulomb} at \SI{150}{\mega\electronvolt} and \SI{14}{\micro\meter} width at the second jet. The simulation window had a size of $(r\times z)=(45\times 440)\cdot k_\mathrm{p}^{-2}$, at a resolution of $\Delta r = \Delta z = 0.033k_\mathrm{p}^{-1}$, with $n_0=6\times 10^{18}\si{\per\cubic\centi\metre}$ inferred from interferometry measurements. In each cell of the mesh, 4 electron and 4 ion macro particles were initialized.

The simulations (see Fig.~\ref{fig:IonChannels}b-e) indeed show a cone-like structure appearing in the ion distribution in the trail of the wake. While our shadowgraphy diagnostic is sensitive to diffraction caused by changes in the local electron density, the ion distribution itself is not visible. However, the plasma wave decays after around $\SI{400}{\micro\meter}$ behind the driver such that the large charge imbalance vanishes and the plasma becomes quasi-neutral, leading to approximately equal electron- and ion-distributions from $\SI{400}{\micro\meter}$ to $\SI{700}{\micro\meter}$. As a result, also the electron distribution exhibits the cone-shaped structure, which allows us to observe this ion motion using shadowgraphy. 

For better comparison with the experimental data, we simulated the propagation of the probe through the electron distribution calculated in the PIC simulation (see the appendix for more information). The synthetic shadowgram, shown in Fig.~\ref{fig:IonChannels}b, is in excellent agreement with the experimental data and reproduces the same diffraction features. The radial velocity of the ion momentum $m_\mathrm{i} v_\perp^{\mathrm{sim}} \sim \SI{4}{\kilo\electronvolt\per c}$ is also compatible with the measured $m_\mathrm{i} v_\perp^{\mathrm{exp}} = \SI[parse-numbers=false]{4.1^{+1.6}_{-1.4}}{\kilo\electronvolt\per c}$.

However, our analysis shows that the mechanism causing the ion motion differs from common ion channel formation due to Coulomb explosion\cite{Mangles:2005wt,Tarkeshian:2018eg}. While a Coulomb explosion leads to a radial expulsion of ions, and hence an annularly-shaped distribution, the ion density in our simulations also increases close to the propagation axis. The reason for this is that the ions in a plasma wave experience radial focusing and defocussing fields in alternation. The net effect of such oscillating forces can be calculated using the ponderomotive formalism. In the non-relativistic limit, which is justified since $v_{\perp}=\SI{0.0017}{c} \ll c$, the ponderomotive force exerted by the plasma wave is\footnote{The derivation of this formula is analogous to ponderomotive motion in a laser field as outlined in common text books, e.g. Ref.\cite{Macchi:2013vx}.}
\begin{equation}
\vec{F}_\mathrm{pond,PW} = -\frac{e^2}{4\omega_\mathrm{p}^2} \vec{\nabla} |\vec E_\mathrm{PW}|^2,
\end{equation}
where $\vec E_\mathrm{PW}$ is the local amplitude vector of the wakefield. In contrast to the well-known ponderomotive force of a laser pulse, the plasma wave amplitude remains almost constant over many periods (equivalent to a flat envelope) so the ponderomotive force of the plasma wave acts mainly radially. Since the radial electric fields of a plasma wave vanish on axis, the intensity gradient points towards $r \rightarrow 0$ for ions close to the symmetry axis, which results in the formation of a density peak on axis and an annular region of ions expanding outwards, as visualized in Fig. \ref{fig:IonChannels}e. This effect was predicted in analytic and numerical studies of laser-driven waves by \citet{Gorbunov:2001cu,Gorbunov:2003bm}, for intense electron beam drivers by \citet{Rosenzweig:2005hf} and for self-modulated plasma wakefield accelerators by \citet{Vieira:2012cz,Vieira:2014de}. 

However, despite the prediction of a similar ponderomotive ion motion for laser-driven plasma waves, we only observed the diffraction pattern behind electron drivers. This observation can be explained by the different field gradients generated by both types of drivers. Electron bunches can self-focus to sizes of or below the skin depth\cite{Keinigs:1987tt}, $\sigma_r \lesssim k_\mathrm{p}^{-1} = \lambda_\mathrm{p}/2\pi$, which leads to strong transverse gradients that in turn cause noticeable ion motion. 
In contrast, \citet{Gorbunov:2001cu} found that the depth and profile of the ion channel for laser-driven plasma waves depends to a large degree on the laser pulse width $w_0$. For laser waist sizes $w_0 \gtrsim \lambda_\mathrm{p}/2 = \pi k_\mathrm{p}^{-1}$, the ion profile resembles a shallow on-axis depression channel, while for smaller laser waists the ion channel becomes deeper and the shape similar to the electron-driven case with a maximum on axis. Only the latter will lead to an electron distribution that can be detected using shadowgraphy, because the diffraction scales with the second derivative of the density. 
In our measurements the plasma wavelength in the first and second jet was $\lambda_\mathrm{p}<\SI{19}{\micro\meter}$ ($n_0 \gtrsim \SI{3e18}{\per\cubic\centi\meter}$). Hence, the laser waist would need to be smaller than $\sim\SI{9.5}{\micro\meter}$, which is well below both the Gaussian waist of \SI{25}{\micro\meter} and the matched spot size $w_0 = 2\sqrt{a_0}k_\mathrm{p}^{-1}$, explaining the missing diffraction feature for the laser-driven case.

We now concentrate on the motion of the outwards expanding ions. The kinetic energy of an ion tends towards the initial ponderomotive potential $\Phi_\text{pond, PW} = (e^2/4m_\mathrm{ion}\omega_\mathrm{p}^2)\cdot |\vec E_\mathrm{PW}|^2$. Hence, the terminal velocity depends on the initial radial position $r_0$ of the ion, cf. momentum vectors in Fig.~\ref{fig:IonChannels}. Ions located further away from the wake's center will only experience a weak ponderomotive force and reach smaller velocities than ions with smaller initial radial position. 
Once the wake depletes and becomes quasi-neutral, the ions move mainly ballistically and the trajectories of ions with different velocities will cross. At this point the amplitude of the transverse density modulation reaches its peak, which also results in a stronger Fresnel diffraction of the probe. However, most of the diffraction signal arises from the border between the low density ring left behind by the ions and the high-density region. The expansion velocity of this ring, which will result in the cone-like shape, is determined by the velocity of the inner-most high-density region. 
Initially those are ions from the central region ($r_0\sim 3 k_\mathrm{p}^{-1}$), but once these ions overtake the slower ions with $r_0\gg 3 k_\mathrm{p}^{-1}$, the cone's shape is determined by these slower ions. We observed this behavior in both experiment and simulations, where the initial opening angle just behind the plasma waves is larger than it is further behind the wakefield. It should be noted that the effect is an import energy dissipation channel in plasma wakefields, as it directly transfers energy from the plasma wave to the ion background.

As mentioned in the introduction, one important feature of plasma wakefield formation is that it scales relative to the plasma parameters, i.e.~with $n_\mathrm{b}/n_0$ and $k_\mathrm{p} \sigma_{z|r}$. Accordingly, most results are scalable to other plasma densities, time and length scales. The high current of laser-accelerated beams generally gives access to higher plasma densities than conventional accelerators. For instance, results from our laser-driven 1-mm-long PWFA operating at \SI[parse-numbers=false]{10^{19}}{\per\cubic\centi\meter} can be scaled to a 10-cm-long PWFA operating at \SI[parse-numbers=false]{10^{15}}{\per\cubic\centi\meter}. Accordingly, our observation of the ion motion persisting up to 50 picoseconds implies that a PWFA with equivalent driver parameters at densities of \SI[parse-numbers=false]{10^{15}}{\per\cubic\centi\meter} would observe ponderomotive ion motion on the time scale of nanoseconds. Hence, our results on ion motion have immediate implications for the design of much larger, low-density PWFAs and the use of bunch trains or self-modulated beams. 

\subsection*{Conclusions and Outlook}

We have used a laser wakefield accelerator, driven by a 100-TW-class laser, to study beam-driven plasma waves and dynamics. By blocking the pump laser of the wakefield accelerator, we could unambiguously show that laser-accelerated electron beams can drive plasma waves at densities of $\sim \SI[parse-numbers=false]{{10}^{19}}{\per\cubic\centi\meter}$. We observe that pre-ionizing the gas target is important in order to effectively drive a plasma wave with bunches having undergone emittance growth in a laser-blocking foil. 

Importantly, the few-cycle shadowgraphy diagnostic not only gives access to femtosecond dynamics of the plasma wakefield, but also allows us to study the electron density evolution over the time scale of picoseconds in a single shot. In doing so, we observed a cone-like diffraction pattern and simulations clearly attribute this feature to ion motion induced by the ponderomotive force of the beam-driven plasma wakefield. As the electron distribution follows the ion motion, the plasma density profile remains perturbed picoseconds behind the plasma wave. This feature is not observed for laser-driven plasma waves, which also allows us to distinguish laser- and beam-driven plasma waves in our experiment.

Due to the physics of PWFA, results obtained at high plasma density using LWFA-electrons can be immediately scaled to low-density scenarios relevant especially for large-scale future PWFA accelerators. The observed ion motion should therefore also occur at longer time scales at conventional PWFA facilities. Indeed, the same feature has been independently observed in recent experiments at the FACET user facility at SLAC\footnote{M. Downer, private communication. Due to the use of Lithium instead of Hydrogen and the much lower plasma density (\SI{5e16}{\per\cubic\centi\meter}), the feature only appears after $\sim\SI{100}{\pico\second}$.}. This demonstrates that compact laser-driven setups can serve as a viable addition or even alternative to large-scale accelerator facilities in beam-driven plasma physics and accelerator research.

In the near future, petawatt laser systems such as the ATLAS-3000 laser in Garching or the Draco-PW laser in Dresden will be able to generate Joule-class (\si{\nano\coulomb} $\times$ \si{\giga\electronvolt}) electron beams\cite{Wang:2013el,Couperus:2017bg}. Using these systems, different regimes of beam-driven wakefield acceleration will be accessible using laboratory-scale systems, e.g.~to produce scaled versions of meter-long PWFAs, bright $\gamma$-ray sources\cite{Ferri:2018ee} or to generate highest-quality electron beams\cite{Hidding:2012ep}, with the latter having the potential to drive compact free electron lasers.

\section*{APPENDIX}
\appendix
\subsection{Experiment configurations}
\begin{table*}
\begin{ruledtabular}
\begin{tabular}{l c c c }
  &
     Experiment 1 & 
     Experiment 2 & 
     Experiment 3 \\
   \hline
  Diameter first jet & 
    \SI{3}{\milli\meter} & 
    \SI{5}{\milli\meter} & 
    \SI{3}{\milli\meter} \\
  Density of first jet & 
    \SI{3.2e18}{\per\cubic\centi\meter} & 
    \SI{2.9e18}{\per\cubic\centi\meter} & 
    \SI{5.6e18}{\per\cubic\centi\meter} \\
  Charge from first jet &
    \SI{200}{\pico\coulomb} & 
    \SI{900}{\pico\coulomb} & 
    \SI{520}{\pico\coulomb} \\ 
  Diameter of second jet & 
    \SI{1}{\milli\meter} & 
    \SI{3}{\milli\meter} & 
    \SI{1}{\milli\meter} \\
  Density of second jet & 
    \SI{1.9e19}{\per\cubic\centi\meter} & 
    \SI{6.0e18}{\per\cubic\centi\meter} & 
    \SI{6.1e18}{\per\cubic\centi\meter} \\
  Jet separation &
    \SI{3}{\milli\meter} &
    \SI{10}{\milli\meter} &
    \SI{3.5}{\milli\meter} \\
  Tape &
    - &
    \SI{15}{\micro\meter} Mylar &
    - \\
  Separation tape to first jet &
    - &
    \SI{2}{\milli\meter} &
    - \\
  Ionizing beam &
    - &
    \SI{60}{\milli\joule} &
    - \\
\end{tabular}
\end{ruledtabular}
\caption{Configurations for Experiment 1,2 and 3. The uncertainties of the densities are \SI{\pm 0.4e18}{\per\cubic\centi\meter}. The charge is measured in the interval between \SIrange{25}{400}{\mega\electronvolt}.}
\label{table:Configuration}
\end{table*}

In this work we have presented three experiments, each with a different target configuration. 
All parameters of the respective setups are summarized in Tab.~\ref{table:Configuration}. We define the respective entrance and exit of the gas jets with the position where the plasma starts becoming visible (corresponding to $\sim \SI{1e17}{\per\cubic\centi\meter}$). The density ramps are \SIrange{0.5}{1}{\milli\meter} long, depending on the nozzle type and if a shock-front is present. The separation between the jets is the length between the exit of the first and entrance of the second jet.
The densities were determined with interferometric measurements and verified with the plasma wavelength from shadowgrams, and the uncertainty is found to be about $\SI{\pm 0.4e18}{\per\cubic\centi\meter}$. Unless otherwise stated these uncertainties apply. 

\subsection{Electron beam spectra and beam profile}

\begin{figure}[h!]
  \centering
  \includegraphics*[width=\columnwidth]{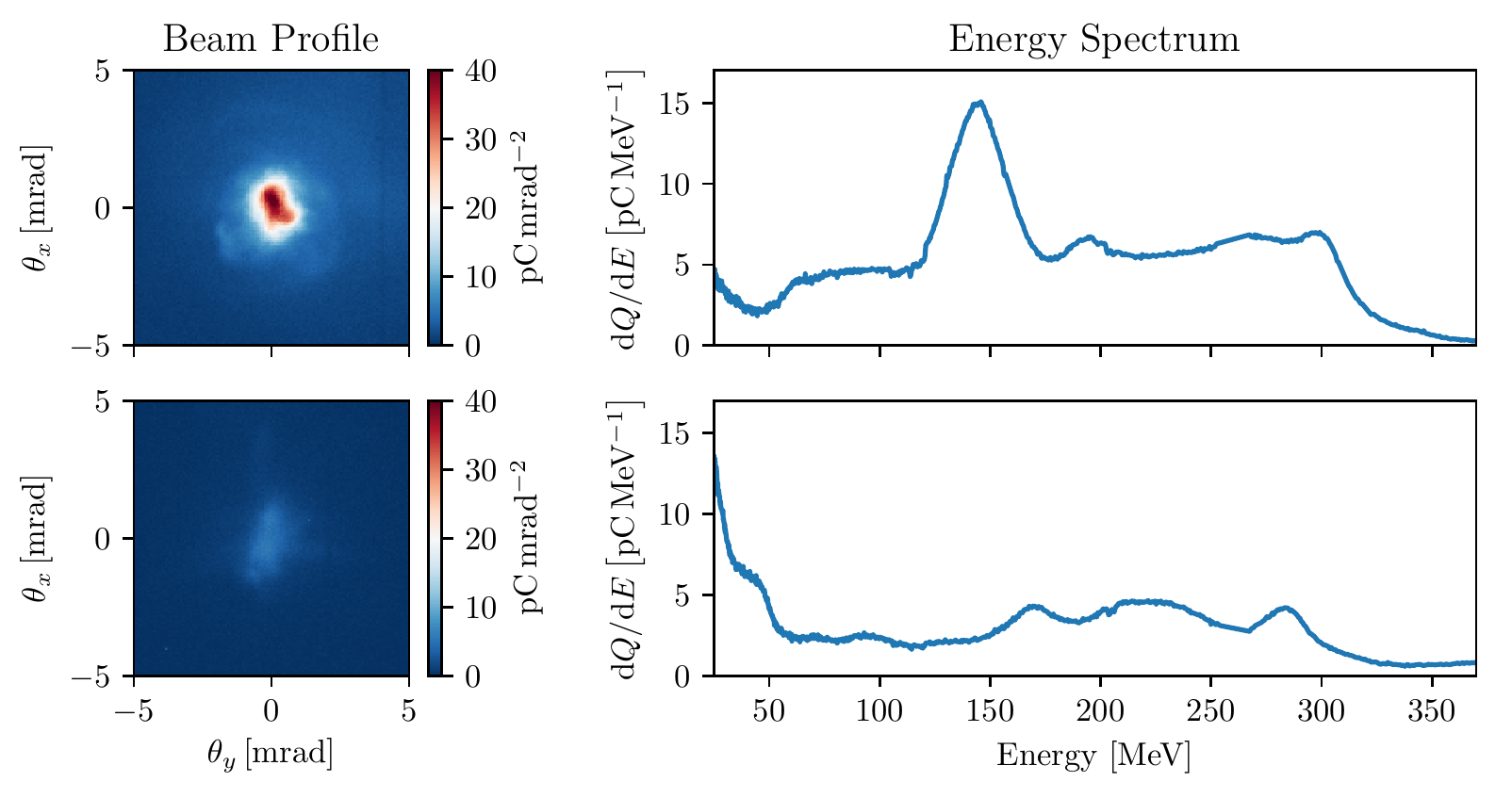}
  \caption{Beam profile and energy spectra of representative shots with only the first gas jet (top) and with tape and second gas jet (bottom).}
  \label{suppl:ElectronSpectra}
\end{figure}

Fig.~\ref{suppl:ElectronSpectra} shows representative electron beam spectra and profiles from experiment 2. The 5 mm long first jet with the shock-front injector was operated at a density of \SI{2.9e18}{\per\cubic\centi\meter}. This resulted in beams with \SI{900}{\pico\coulomb} charge, spectra as representatively shown in Fig. \ref{suppl:ElectronSpectra} and \SI{1.7}{\milli\radian} FWHM divergence. The beam charge was characterized using an absolutely calibrated scintillating screen, see \citet{Kurz:2018ji} Note that in contrast to prior work, the shock-front injector was operated with optimized beam charge and divergence, which results in a broad energy spectrum. Also, while the second jet clearly affected the spectrum and divergence of the electron beam, we did not observe any clear acceleration or deceleration effect. This is mainly due to the shot-to-shot fluctuations and the above-mentioned beam energy spread. The charge detected in the spectrometer decreased from \SI{900}{\pico\coulomb} to \SI{550}{\pico\coulomb} when the tape was inserted and the second jet was activated, a reduction similar to \citet{Chou:2016ei}

\subsection{Few-cycle pulse generation}

The probe beam was derived from the main ATLAS beam using a half-inch mirror. This beam was then guided through a \SI{1}{\milli\meter} thick fused silica window to a probe table outside the vacuum target chamber. The diameter and energy were adjusted using an iris and ND filters to about $\SI{8}{\milli \metre}$ and $\SI{1}{\milli\joule}$, respectively. A dispersive mirror array together with a variable-thickness glass wedge pair compensated the group delay dispersion (GDD) accumulated during pre-fiber propagation and therefore ensured effective self-phase modulation (SPM) inside the Argon-filled hollow core fiber with an inner-diameter of \SI{240}{\micro \metre} and a length of \SI{0.9}{\metre}. With an Argon pressure of \SI{500}{\milli \bar}, about \SI{400}{\micro \J} were transmitted though the fiber. Thereafter, a second array of dispersive mirrors and a wedge pair were used to compress the pulse close to its Fourier limit.

\subsection{Simulated shadowgrams}
Previous studies on FCS have used 3D-cartesian PIC simulations with a separately initialized probe beam to simulate shadowgrams \cite{Siminos:2016gh}. However, this approach becomes impractical for the large simulation windows as required in our case. Instead, we calculated a qualitative approximation of the shadowgrams of the simulated ion channels from quasi-3D simulation data in post-processing. Using the dispersion relation of a cold plasma, we used the electron distribution to calculate the phase shift of a plane monochromatic wave traveling perpendicularly through the moving plasma in a static approximation. The electron distribution of the radially symmetric simulation was mapped onto a 3D grid where $\vec{e}_z$ is the direction of propagation of the driver and $\vec{e}_y$ the direction of propagation of the probe.
Each layer in the $\vec{e}_{x|y}$ plane was shifted by $c\Delta y$ in $\vec{e}_z$ direction, such that the distribution appeared as moving with the speed of light as the probe propagated through it. 

While our results show good agreement with the shadowgrams observed in experiment, it should be noted that there are a few limitations to our approach. First, it is only valid if the plasma wave does not evolve significantly while the probe transverses it. This is usually the case in wakefield acceleration and for all situations treated in this study, but special cases such as wave evolution in density gradients would be an exception. Here one would need to use simulation data from different time steps.
Furthermore, the cold plasma approximation is strictly only valid behind the plasma wave. Within the plasma wave the diffraction can be overestimated due to the the reduced refractive index of relativistic electrons. If needed, this could be solved by analyzing not only density maps, but the test particle data including their momenta.

\subsection*{Acknowledgements}

The authors thank J.\ Vieira (IST) and M.\ Downer (U. Texas) for helpful discussions.\\
	
This work was supported by DFG through the Cluster of Excellence Munich-Centre for Advanced Photonics (MAP EXC 158), TR-18 funding schemes, by EURATOM-IPP, and the Max Planck Society. The authors gratefully acknowledge the Gauss Centre for Supercomputing e.V. (www.gausscentre.eu) for funding this project by providing computing time on the GCS Supercomputer SuperMUC at Leibniz Supercomputing Centre (www.lrz.de) under project id pn69ri. The authors also would like to acknowledge the \textsc{osiris} Consortium,  consisting of University of California, Los Angeles (USA) and Instituto Superior T\'enico (Lisbon, Portugal) for the use of \textsc{osiris} and the visXD framework. S.M.H. was supported by a visitor grant from the Center for Advanced Studies (CAS) at LMU Munich.\\[2mm]

\subsection*{Author contributions}
M.F.G., H.D., A.D., J.G., S.S., G.S., S.M.H. and S.K. set up and/or performed the experiment. H.D. set up the few-cycle probe. M.F.G. analyzed the data and performed simulations. All authors discussed the results. M.F.G. and A.D. wrote the paper. S.K. supervised the project.

\bibliography{references.bib}

\onecolumngrid
\newpage

\begin{center}
\textbf{\Large Direct observation of plasma waves and dynamics \\induced by laser-accelerated electron beams}\\[5mm]

M. F. Gilljohann, H. Ding, A. D\"opp et al.\\[5mm]

\textbf{\large Supplemental Materials}
\end{center}

\setcounter{equation}{0}
\setcounter{figure}{0}
\setcounter{table}{0}
\makeatletter
\renewcommand{\theequation}{S\arabic{equation}}
\renewcommand{\thefigure}{S\arabic{figure}}
\renewcommand{\thetable}{S\arabic{table}}

\section*{Differences between PWFA and LWFA}

As discussed in the introduction, plasma wakefield are formed by either the space-charge of a particle bunch or the ponderomotive force of a laser pulse. For  readers not familiar with the differences between both schemes, we discuss some basic differences regarding plasma wakefield generation and applications to particle acceleration. 

In the linear regime of wakefield acceleration, the plasma wave perturbation is given by
$$
\left( \frac{\partial^2}{\partial t^2} + \omega_\mathrm{p}^2 \right)\frac{n_\mathrm{e}}{n_0}=\underbrace{-\omega_\mathrm{p}^2\frac{n_\mathrm{b}}{n_0}}_{\text{Space charge}} + \underbrace{c^2\Delta \frac{a^2}{2}}_{\text{Ponderomotive force}}  ,
$$
where $\vec{a}=e\vec{A}/m_\mathrm{e}c^2$ is the normalized vector potential. Hence, both the space charge term $n_\mathrm{b}/n_0$ and ponderomotive force ${a^2}/{2}$ have similar influence on the plasma wave formation. Depending on the amplitude of these terms, either a linear plasma wake is generated or a highly non-linear ion-cavity forms. In the linear regime, both the space charge and ponderomotive force produce only a density perturbation ($n_\mathrm{b}/n_0\ll 1$, $a_0\ll1$). Instead if a beam driver reaches $n_\mathrm{b}/n_0>1$, $k_\mathrm{p}\sigma_z<1$, $k_\mathrm{p}\sigma_r<1$ or a laser $a_0\gg1$, the non-linear blow-out regime is reached.

However, while the shape and amplitude of beam and laser-driven wakefields are very similar, there are some important differences between both cases.

\begin{itemize}
\item First, in the case of wake excitation via space-charge, plasma electrons outside the bunch radius $\sigma_r$ are still affected by the Lorentz force of the electron beam, which leads to a typical transverse extent of up to $k_\mathrm{p}^{-1}$. Also, according to Gauss' law, the fields for plasma electrons outside of the bunch do not depend on the actual density distribution within the beam. For laser-driven wakes, the plasma wave formation is caused by the ponderomotive force and hence, the plasma wave 
approximately extends out to the waist $w_0$. But since the laser tends to evolve towards the matched spot size at a given density, the transverse size of laser driven wakes can be much larger than for beam-driven wakes.
\item Furthermore, as the space charge fields are unipolar in contrast to the oscillating laser fields, a beam driver can generate similar wakefields to a laser driver at orders of magnitude lower field amplitudes. An immediate consequence of this is that particle beams for wakefield acceleration do not ionize neutral gas as efficiently as lasers. On one hand this creates the need for pre-ionization schemes, while on the other hand, it also allows for advanced injection schemes based on selective ionization.
\item Last, relativistic electron beams propagate at a velocity $\beta_\mathrm{e} = v_\mathrm{e}/c = 1/\sqrt{1-\gamma^2}$, which is very close to the speed of light $c$ in the case $\gamma \gg 1 $. In contrast, laser beams propagate at a group velocity $\beta_\text{laser}=\sqrt{1-\omega_\mathrm{p}^2/\omega_0^2}$, where $\omega_0$ is the angular frequency of the laser. As $\beta_\text{laser}<\beta_\mathrm{e}$ in most cases, electron beams slowly advance with respect to the laser, which leads to dephasing. This limits the maximal energy gain in laser wakefield accelerators. While some density tailoring schemes have been proposed to mitigate this problem, a TeV-scale laser wakefield accelerator would almost certainly require multiple acceleration stages. In contrast, the velocity difference between the driver and the so-called witness beam in electron beam-driven PWFA is negligible, i.e. there is no dephasing and energy gain is limited by the transformer ratio. The higher phase velocity of the wakefield also prohibits self-injection of background plasma electrons, which makes this scheme free of dark current.
\end{itemize}
\newpage

\subsection*{Full field of view shadowgrams for different configurations}

\begin{figure}[h!]
  \centering
  \includegraphics*[width=\columnwidth]{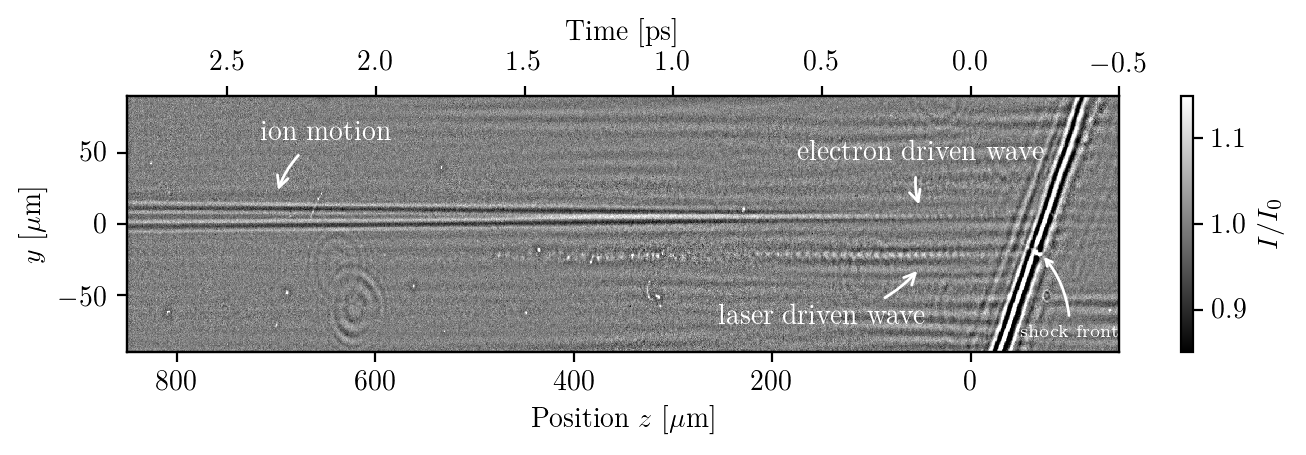}
  \caption{Full view of the shadowgrams from Fig.\ref{fig:LWFA_PWFA}. Note that there is no cone-like diffraction pattern in the trail of the laser driven plasma wave. Furthermore, we observe a bright spot at the shock-front, located at the height of the laser-driven wave, but missing at the height of the electron-driven wave. This emission is wavebreaking radiation\cite{Thomas:2007cc} and its absence in the electron-driven case indicates that the longitudinal wakefields are weaker than in the laser-driven case.}
  \label{suppl:LWFA_PWFA}
\end{figure}

\begin{figure}[h!]
  \centering
  \includegraphics*[width=\columnwidth]{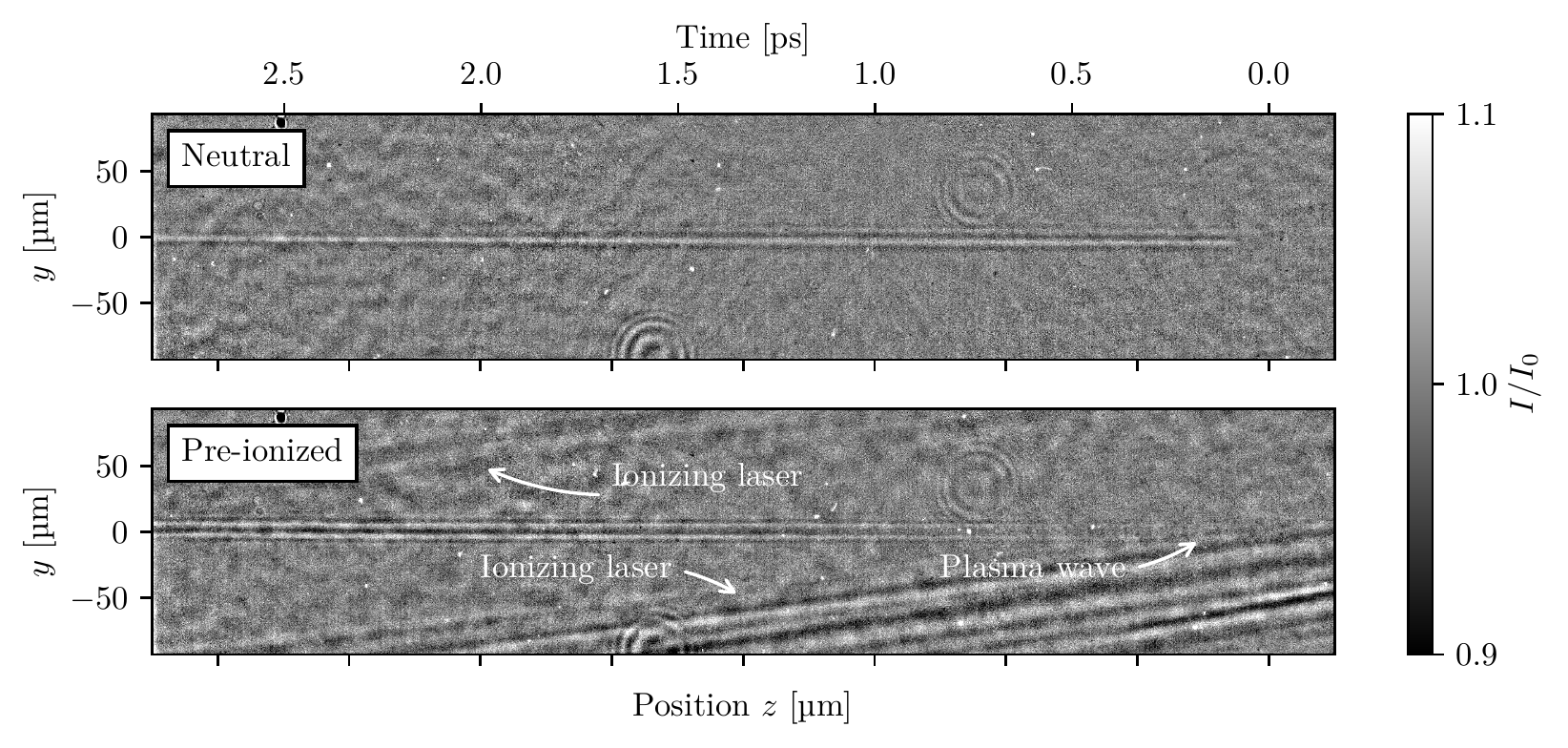}
  \caption{Full view of the shadowgrams from Fig.~\ref{fig:PreionizedNeutral}, with subtracted background and flatfield correction. The beam propagates from the left to the right, while the ionizing beam propagates from the top right to the left at an angle of \SI{173}{\degree}. Note that in contrast to Fig.~\ref{fig:PreionizedNeutral} in the main text, here we only apply background correction for the drive beam and not the ionizing pulse in order to illustrate the geometry.}
  \label{suppl:PWFA}
\end{figure}

\newpage
\subsection*{More simulation plots}
\begin{figure}[h!]
  \centering
  \includegraphics*[width=\columnwidth]{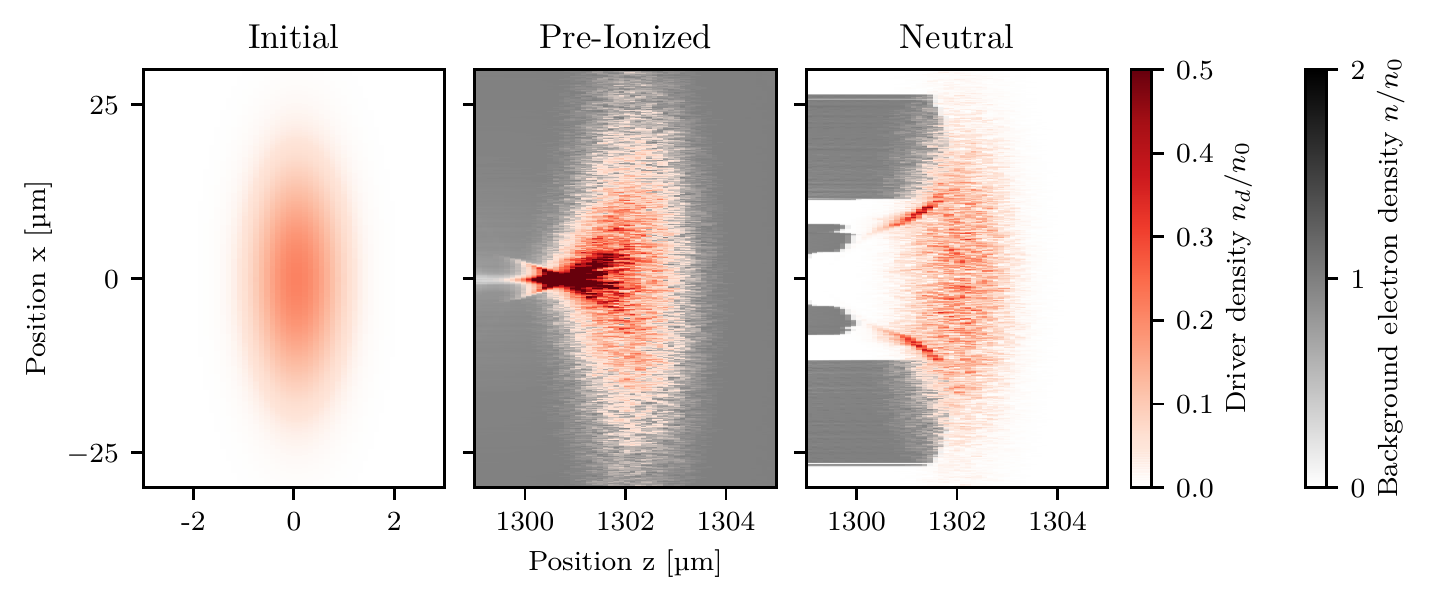}
  \caption{Driver distributions of the simulations in Fig~\ref{fig:PreionizedNeutral} before the interaction (\textit{left}) and with pre-ionized (\textit{middle}) and neutral gas (\textit{right}) after \SI{1.3}{\milli\meter} of propagation. It is clear that the tail of the electron beam is well-focused in the pre-ionized case, where the beam sent into a neutral has undergoes only weak focusing and only partially ionizes the plasma.}
  \label{suppl:Drivers}
\end{figure}

\begin{figure}[h!]
  \centering
  \includegraphics*[width=\columnwidth]{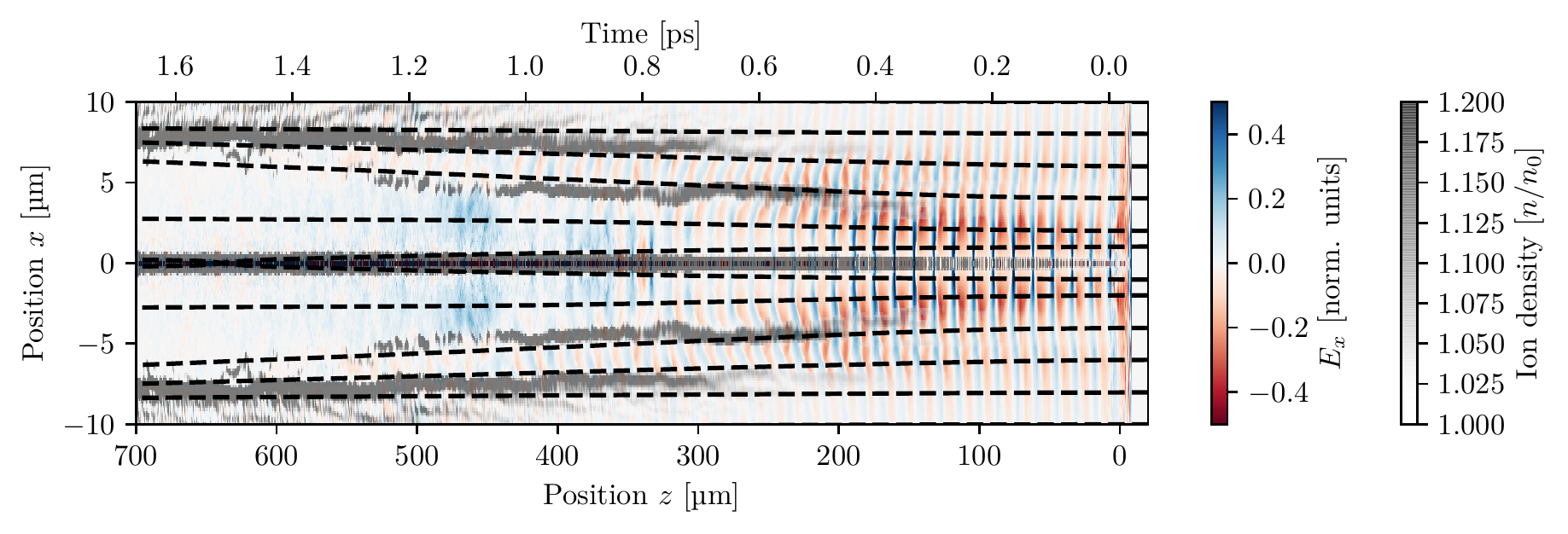}
  \caption{Large view of the simulated transverse electric field and ion density from Fig.4. Dashed lines show selected ion trajectories.}
  \label{suppl:Ponderomotive}
\end{figure}

\end{document}